\documentclass[a4paper,11pt]{article}
\usepackage{jinstpub} 
\usepackage{lineno}

\title{\boldmath Lepton Flavour Universality tests in $b \to c l \nu$ decays at LHCb}

\collaboration[c]{on behalf of the LHCb collaboration}

\author{R. Mohammed}
\affiliation{Department of Physics, University of Oxford,\\
Parks Road, Oxford, United Kingdom}

\emailAdd{rizwaan.mohammed@physics.ox.ac.uk}

\abstract{The Standard Model predicts that the electroweak couplings to the three charged leptons are identical. However, in the last decade, experimental measurements have suggested that semileptonic processes involving taus could have a slightly enhanced decay rate compared to their muonic counterparts. If confirmed, this would be an unambiguous sign of New Physics, with various scenarios introducing additional interactions that couple preferentially to the third generation. Two recent lepton universality tests performed at LHCb are presented in these proceedings of the first edition of the New Frontiers in Lepton Flavour workshop.}

\begin{document}
\maketitle
\flushbottom

\section{Introduction}
\label{sec:intro}

Lepton Flavour Universality (LFU) is an accidental symmetry of the Standard Model (SM); the symmetry is not caused by any underlying conservation law. It predicts that the electroweak coupling to each generation of leptons is identical. Consequently, any differences in interactions involving the three charged leptons is solely due to phase-space effects arising from their different masses.

A common way to test LFU is to measure branching fraction ratios with $b \to c l \nu$ decays, i.e.
\begin{equation}
    \mathcal{R}(H_c) = \frac{\mathcal{B}(H_b \to H_c \tau \nu_\tau)}{\mathcal{B}(H_b \to H_c \mu \nu_\mu)} \; ,
    \label{eq:R_Hc}
\end{equation}
where $H_b$ and $H_c$ are beauty and charm hadrons, respectively. These measurements are powerful tests of LFU, since the ratio allows cancellation of some theoretical and experimental uncertainties.

Two recent tests of LFU performed at LHCb are presented here. The first is a simultaneous measurement of $\mathcal{R}(D^0)$ and $\mathcal{R}(D^*)$ with LHCb Run 1 data. The second is a measurement of $\mathcal{R}(D^*)$ with partial LHCb Run 2 data.
\section{\boldmath $\mathcal{R}(D^0) - \mathcal{R}(D^*)$ muonic} \label{sec:RDst_muonic}
This measurement, reported in  Ref. \cite{RDz_Rdst_muonic}, involves the reconstruction of the $\tau$ in the muonic decay mode, meaning that the $\mu$ and $\tau$ decays in eq. \eqref{eq:R_Hc} have the same visible final state. This allows $\mathcal{R}(D^0)$ and $\mathcal{R}(D^*)$ to be extracted directly without the need for an external normalisation. Previously, LHCb measured $\mathcal{R}(D^*)$ with this final state using the Run 1 dataset, as reported in Ref.~\cite{Rdst_muonic}. The new analysis extends the previous work by making a simultaneous measurement of $\mathcal{R}(D^0)$ and $\mathcal{R}(D^*)$ with the same dataset.

The LHCb Run 1 dataset corresponds to an integrated luminosity of 3 $\textrm{fb}^{-1}$ collected at a centre of mass energy of 7 and 8 TeV. The dataset is first split into two samples, one of which has enhanced $D^{*+}\mu^-$ combinations, and the other has enhanced $D^{0}\mu^-$ as the $D^{*+}$ state is vetoed. Of the two samples, the $D^{0}\mu^-$ is approximately five times larger due to its higher branching fraction and better efficiency.

As with any semileptonic study at a hadron collider, the presence of unreconstructed neutrinos in the final state compromises the measurement as the $B$ rest frame cannot be reconstructed exactly. Instead, it must be approximated, which is done by assuming the proper velocity of the $B$ is equal to the proper velocity of the visible component $(D^{(*)}\mu)$ along the $z$ axis. This gives the $z$ component of the $B$ momentum, then the other components can be determined from knowledge of the $B$ flight direction. With this estimate, the rest frame quantities of interest can be calculated. These are: the invariant mass of the lepton-neutrino system, $q^2 = (p_B - p_{D^{(*)}})^2$; the squared missing mass, $m_{miss}^2 = (p_B - p_{D^{(*)}} - p_\mu)^2$; and the muon energy in the $B$ rest frame, $E_\mu ^*$; where $p_B$, $p_{D^{(*)}}$ and $p_\mu$ are the four-momenta of the $B$, $D^{(*)}$ and the muon respectively.

Track isolation is used to reject background from processes with additional tracks. This aims to isolate the signal candidate from the rest of the event. The isolation is performed using a boosted decision tree (BDT), which predicts whether a given track is compatible with the $B$ vertex. This BDT efficiently reduces contributions from $B \to D^{**} \mu \nu$ processes. 

In addition, the BDT allows a background-enhanced regions of the data to be selected; these are used as control regions to give an improved background modelling. Three separate control regions are used for both the $D^0 \mu^-$ and the $D^{*+} \mu^-$ datasets. The first region contains an extra pion track, used to constrain light excited $D$ mesons, which decay in the form $B \to (D^{**} \to D^* \pi)\, l \nu$. Similarly, the second control region uses two additional pion tracks to model heavier excited $D$ mesons, which decay in the form $B \to (D^{**} \to D^* \pi \pi)\, l \nu$. The third control region contains at least one additional kaon track, which models $B \to D^{(*)}DX$ backgrounds. These backgrounds are all modelled using simulations.

Other background processes considered in this study include decays of the form $B \to D^{(*)}hX$, where $h$ is a charged hadron that is misidentified as a muon. A data-driven method is used to estimate the rate of this process and to model it in the final fit. Also, there are ``combinatorial'' backgrounds, which are random combinations of $B$ and $D^{(*)}$ daughters leading to fake $B$ and $D^{(*)}$ candidates. These are modelled from data by combining same-sign particles into fake candidates.

\begin{figure}
    \centering
    \includegraphics[width=0.87\textwidth]{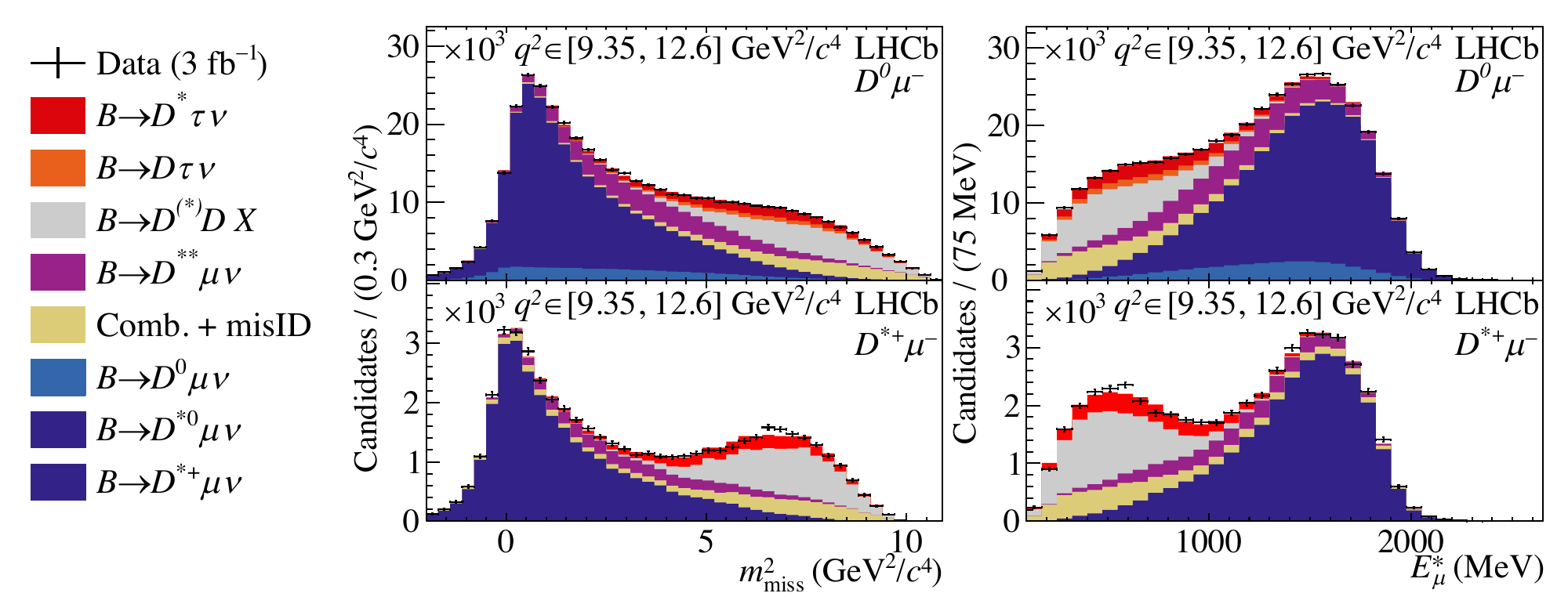}
    \caption{Fit projections for $m_{miss}^2$ and $E_\mu ^*$ in the highest $q^2$ bin for the $D^0 \mu^-$ sample (top) and the $D^{*+} \mu^-$ sample (bottom).}
    \label{fig:RDst_muonic_fits}
\end{figure}

To extract the parameters of interest, a three-dimensional maximum likelihood template fit is carried out. The fit variables are the rest frame quantities: $q^2$, $m_{miss}^2$ and $E_l ^*$. The fit uses eight simultaneous data samples (the signal and three control regions for the $D^0 \mu^-$ and the $D^{*+} \mu^-$ datasets). Furthermore, this study developed two separate fitting algorithms. The first method fitted all signal and control regions simultaneously. The second method fitted the control regions to obtain corrections for the most signal-like backgrounds, then used these results to constrain parameters in a fit to the signal samples. The agreement between the two was checked extensively. The fit projections in the highest $q^2$ bin are shown in Figure~\ref{fig:RDst_muonic_fits}.

The results of the measurement are:
\begin{equation}
    \mathcal{R}(D^{*+}) = 0.281 \pm 0.018\, \mathrm{(stat.)} \pm 0.024\, \mathrm{(syst.)} \;,
\end{equation}
\begin{equation}
    \mathcal{R}(D^0) = 0.441 \pm 0.060\, \mathrm{(stat.)} \pm 0.066\, \mathrm{(syst.)} \;,
\end{equation}
where the first uncertainty is statistical and the second is systematic. The dominant systematic uncertainties arise from the limited size of the simulation samples used to create the fit templates, the uncertainty of the $B \to D^{(*)}DX$ template shape, and the uncertainty from form factor parameters describing the signal and the $B \to D^{**} \mu \nu$ background. Overall the result has a $1.9 \sigma$ agreement with the SM~\cite{HFLAV}, and the correlation between $\mathcal{R}(D^{*+})$ and $\mathcal{R}(D^0)$ is $-0.43$. This is the first simultaneous measurement of $\mathcal{R}(D^{0})$ and $\mathcal{R}(D^*)$ at a hadron collider.
\section{\boldmath $\mathcal{R}(D^*)$ hadronic} \label{sec:RDst_hadronic}
The second measurement presented here is a determination of $\mathcal{R}(D^{*})$, where the $\tau$ is reconstructed in the $\tau^+ \to \pi^+ \pi^- \pi^+ (\pi^0)$ final state~\cite{RDst_hadronic_Run2}. This is an update to a previous LHCb measurement \cite{RDst_hadronic_Run1} which performed the same procedure on the Run 1 LHCb dataset. The previous measurement had an integrated luminosity of 3 $\mathrm{fb}^{-1}$, and the new result uses data taken in 2015 and 2016, corresponding to 2 $\mathrm{fb}^{-1}$. Despite the lower luminosity, the new analysis has approximately 40\% more candidates than the previous measurement. This is due to a higher centre of mass energy in Run 2, as well as improvements in the LHCb trigger system.

In contrast to the measurement presented in Section \ref{sec:RDst_muonic}, a separate normalisation channel is employed $(B^0 \to D^{*-} 3\pi^\pm)$, which was chosen as it has the same final state as the signal decay. The analysis measures the ratio $\mathcal{K}(D^*)$, defined as:
\begin{equation}
    \mathcal{K}(D^*) = \frac{\mathcal{B}(B^0 \to D^{*-}\tau^+\nu_\tau)}{\mathcal{B}(B^0 \to D^{*-} \pi^+ \pi^- \pi^+)}\; .
    \label{eq:KDst}
\end{equation}
This can then be converted to a value of $\mathcal{R}(D^{*})$ using:
\begin{equation}
    \mathcal{R}(D^{*}) = \mathcal{K}(D^*)\, \frac{\mathcal{B}(B^0 \to D^{*-} \pi^+ \pi^- \pi^+)}{\mathcal{B}(B^0 \to D^{*-}\mu^+\nu_\mu)}\; .
    \label{eq:RDst_KDst}
\end{equation}
The value of the second ratio in the above equation is an external input, taken from the PDG \cite{PDG}.

The data used for this analysis contains a large background contribution from ``prompt'' $B \to~D^* 3\pi X$ decays. This is reduced using information from the $3\pi$ vertex. In the signal mode, the $3\pi$ system is produced via the intermediate $\tau$ decay, whereas in the background, it comes directly from the $B$ vertex. Variables related to the separation of the $3\pi$ and $B$ vertices are used to train a BDT, which rejects over 99\% of the background at the chosen working point.

Furthermore, there is a significant contribution from $B \to D^{*-}D_s ^+\, (\to 3\pi X)\,X$ events (double charm background). These events mimic the signal topology, as the $D_s$ meson behaves similarly to the $\tau$ in the signal decay. A dedicated BDT was trained using $D_s$ kinematic variables and isolation variables. The BDT reduces the double charm contribution, and is also used as a variable in the signal fit. To improve the modelling of the double charm decays, a separate fit is performed to measure the various production fractions of these modes. These values are then used to constrain the fractions in the signal fit. In addition, the $D_s$ BDT is used to produce a data sample enriched in double charm events. A fit to the pion mass variables is carried out using this control sample to measure the fractions of $D_s \to 3\pi X$ decays. The results of this fit are used to correct the decay fractions in the simulation that makes up the background template in the final signal fit.

\begin{figure}
    \centering
    \includegraphics[width=0.85\textwidth]{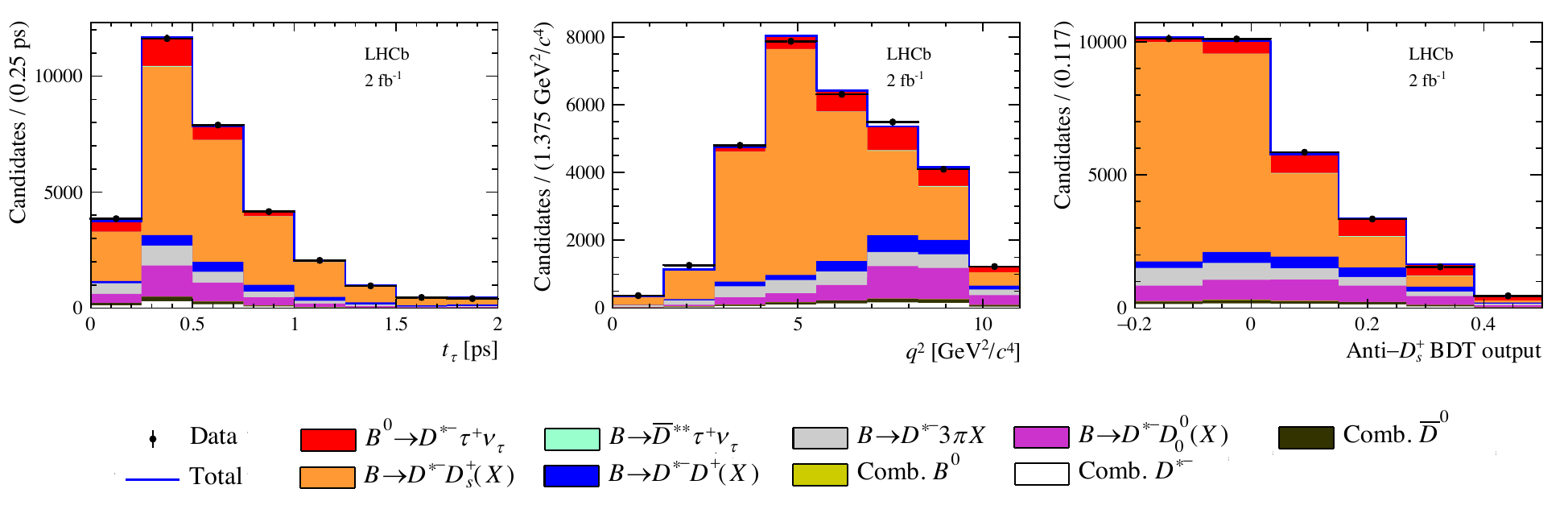}
    \caption{One-dimensional projections of the signal fit used to measure the yield of $B^0 \to D^{*-} \tau^+ \nu_\tau$ events}
    \label{fig:RDst_hadronic}
\end{figure}

Similarly to the analysis presented in Section \ref{sec:RDst_muonic}, a three dimensional maximum likelihood template fit is used in this work. In this case, the fit variables are $q^2$, the $D_s$ BDT, and the $\tau$ lifetime. This signal fit measures the number of $B^0 \to D^{*-} \tau^+ \nu_\tau$ events, and twice as many signal events were found compared to the Run 1 analysis. The signal fit projections are shown in Figure~\ref{fig:RDst_hadronic}. The yield of $B^0 \to D^{*-}3\pi$ is measured from a separate normalisation fit. This is a simple one dimensional fit to the $(D^{*-}3\pi)$ mass, which has a tiny combinatorial background contribution. From this, $\mathcal{K}(D^*)$ is calculated by taking the ratio of signal and normalisation events, correcting for the different efficiencies, and dividing by the $\tau^+ \to 3\pi (\pi^0)\bar{\nu}_\tau$ branching fraction. The measured value is:
\begin{equation}
    \mathcal{K}(D^*) = 1.700 \pm 0.101\, \mathrm{(stat.)} \, ^{\,+0.105}_{-0.100} \, \mathrm{(syst.)} \; ,
\end{equation}
where the first uncertainty is statistical and the second is systematic. The dominant sources of systematic uncertainty are the limited sizes of the simulation samples and incomplete knowledge of the background shapes. From this, $\mathcal{R}(D^*)$ is calculated as:
\begin{equation}
    \mathcal{R}(D^*) = 0.247 \pm 0.015 \, \mathrm{(stat.)} \pm 0.015 \,  \mathrm{(syst.)} \pm 0.012 \, \mathrm{(ext.)}\; ,
\end{equation}
where the final uncertainty comes from the external branching fraction in Eq.~\eqref{eq:RDst_KDst}. When this result is combined with the value obtained in the Run 1 analysis, $\mathcal{R}(D^*)$ is found to be:
\begin{equation}
    \mathcal{R}(D^*) = 0.257 \pm 0.012 \, \mathrm{(stat.)} \pm 0.014 \,  \mathrm{(syst.)} \pm 0.012 \, \mathrm{(ext.)}\; .
\end{equation}
The combined value is consistent with the SM prediction within $1\sigma$~\cite{HFLAV}.
\section{Conclusions and prospects}
\begin{figure}
    \centering
    \includegraphics[width=0.7\textwidth]{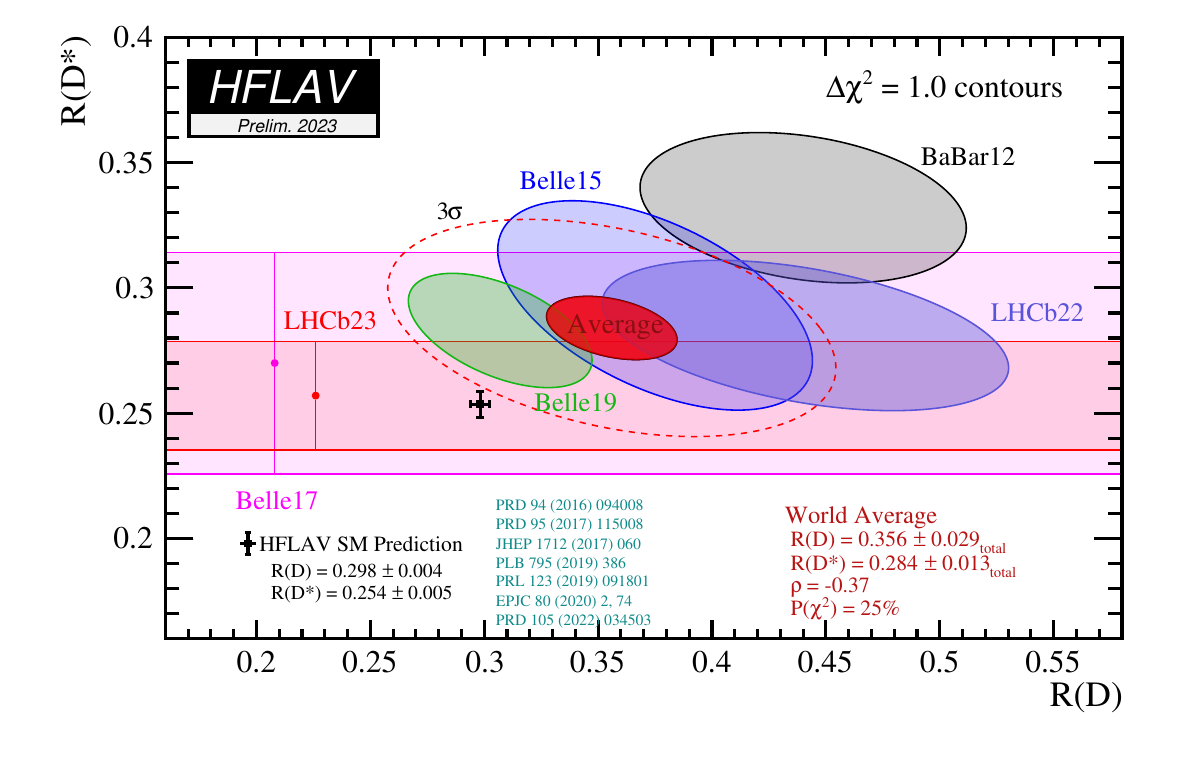}
    \caption{Average experimental values for $\mathcal{R}(D^0)$ and $\mathcal{R}(D^*)$ (red ellipse) with their SM predicted values (black cross)~\cite{HFLAV}}
    \label{fig:HFLAV}
\end{figure}

Previous measurements of $\mathcal{R}(D^0)$ and $\mathcal{R}(D^*)$ have shown tensions with the SM. The world average values for $\mathcal{R}(D^0)$ and $\mathcal{R}(D^*)$, including results from LHCb, BaBar and Belle had a combined 3.3$\sigma$ deviation from the SM. The two new results presented here both have good agreement with the SM, so the tension between the experimental and predicted values is reduced to 3.2$\sigma$ when they are included (see Figure \ref{fig:HFLAV}). 

Further measurements of $R(D^{(*)})$ and other LFU ratios present exciting prospects for potential New Physics (NP) in the lepton sector. The full Run 2 dataset at LHCb is yet to be exploited, and data taking for Run 3 has recently begun. Moreover, complementary analyses from Belle II will be crucial in providing robust measurements of these ratios. Finally, LHCb is also working on angular analyses of $B \to D^* l \nu$ decays using the method presented in Ref.~\cite{angular}. These are potentially more powerful tests of NP as they provide additional observables that can distinguish between NP scenarios and can be model independent.

\bibliographystyle{JHEP}
\bibliography{biblio.bib}


\end{document}